%% file: DPF2019_pinunuLNV_final.tex
\documentclass[11pt]{article}
\usepackage[margin=1in]{geometry}
\pdfoutput=1
\usepackage{graphicx}
\usepackage{wrapfig}
\usepackage{hyperref}

\input econfmacros.tex 

\def\Title#1{\begin{center} {\Large {\bf #1} } \end{center}}
\def\Author#1{\begin{center} {\normalsize {\sc #1} } \end{center}}
\def\Institution#1{\begin{center} {\normalsize {\it #1} } \end{center}}
\def\Abstract#1{\noindent {\normalsize {\bf Abstract:} {\normalfont #1}}}
\def\Conference{\vspace{4mm}\begin{raggedright} {\normalsize {\it Talk presented at the 2019 Meeting of the Division of Particles and Fields of the American Physical Society (DPF2019), July 29--August 2, 2019, Northeastern University, Boston, C1907293.} } \end{raggedright}\vspace{4mm}}

\begin{document}

%
%

\Title{Physics beyond SM with kaons from NA62}

\Author{Roberta Volpe
  \footnote{On behalf of the NA62 Collaboration: R. Aliberti, F. Ambrosino, R. Ammendola, B. Angelucci, A. Antonelli,
    G. Anzivino, R. Arcidiacono, T. Bache, M. Barbanera,
    J. Bernhard
% ...
}}

\Institution{Centre for Cosmology, Particle Physics and Phenomenology \\ Universit\`{e} Catholique de Louvain, Louvain La Neuve, Belgium}

\Abstract{The decay $K^+ \to \pi^+ \nu \bar{\nu}$, with a very precisely predicted branching ratio of less than $10^{-10}$,
  is one of the best candidates to reveal indirect effects of new physics at the highest mass scales.
  The NA62 experiment at the CERN SPS is designed to measure the branching ratio of the $K^+ \to \pi^+ \nu \bar{\nu}$
  with a decay-in-flight technique.
  NA62 took data so far in 2016-2018. Statistics collected in 2016 allowed NA62 to reach the Standard Model
  sensitivity for $K^+ \to \pi^+ \nu \bar{\nu}$, entering the domain of $10^{-10}$ single event sensitivity
  and showing the proof of principle of the experiment. Thanks to the statistics collected in 2017, NA62 surpasses the present best sensitivity.
  The analysis strategy is reviewed and the preliminary result from the 2017 data set is presented.
  A large sample of charged kaon decays into final states with multiple charged particles was collected in 2016-2018.
  The sensitivity to a range of lepton flavor and lepton number violating kaon decays provided by this data set improves
  over the previously reported measurements. Results from the searches for these processes with a partial NA62 data sample are presented.}

\Conference

%
%

\section{Introduction}
The NA62 Experiment \cite{ref:NA62TD} has been designed and built in order to measure
the branching ratio of the decay $K^+ \to \pi^+ \nu \bar{\nu}$.
The importance
of this measurement lies on the extremely theoretical cleanness of this process environment.
This decay is a flavor changing neutral current process which proceeds through box and penguin diagrams,
the quark level amplitude is dominated by the top quark term while the light quarks contributions are suppressed by the GIM mechanism,
and therefore it does not suffer much from hadronic uncertainties.
Within the Standard Model (SM) it is very suppressed and its branching ratio,
computed considering the experimental uncertainties on the CKM angles and CPV phase,
is
\begin{equation}
  BR(K^+ \to \pi^+ \nu \bar{\nu}) = (8.4 \pm 1.0)\times10^{-11} ,
\end{equation}
while, if the SM parameters are taken to be exact, it is $(9.1\pm0.7)\times10^{-11}$ \cite{ref:Buras}.
Beyond standard model scenarios with new sources of flavor violations
foresee large deviation from such SM values, 
for example \cite{ref:Blanke}\cite{ref:Blazek}\cite{ref:Isidori}\cite{ref:Buras2016}\cite{ref:Isidori2}, making this decay an appealing
probe of new physics.
The only measurement of this branching ratio was obtained by E949 collaboration \cite{ref:E949} using the stopping kaon technique and results in:
\begin{equation}
    BR(K^+ \to \pi^+ \nu \bar{\nu}) = (17.3^{+11.5}_{-10.5}) \times 10^{-11}
\end{equation}
Comparing the precision obtained for the SM computation with the experimental results,
it is clear that a more precise measurement is needed and the NA62 experiment has been designed for this purpose.
The NA62 apparatus is described in Sec.\ref{sec:na62}, while the $K^+ \to \pi^+ \nu \bar{\nu}$ analysis strategy
is summarized in Sec.\ref{sec:pinunu}.
The NA62 experiment can be exploited also to search for other deviations from the SM predictions in kaon decays.
Processes which violate the conservation of the total lepton number $L$ (LNV) and of lepton flavor numbers $L_l$, with $l=e,\mu,\tau$, (LFV)
have never been observed.
Yet, they are not forbidden by SM principles.
Neutrino oscillations and hence neutrino masses,
indicate that the lepton family number is not an exact symmetry.
The existence of Majorana neutrino would violate total lepton number $L$ by $|\Delta L| = 2$ units \cite{ref:LNV, ref:LNV2}.
NA62 can search for the three tracks LFV/LNV kaon decays 
with dedicated multi-track trigger streams.
Sec.\ref{sec:LNV} reports the results obtained in the searches for $K^+ \to \pi^- l^+ l^+$, $l=e,\mu$, with a subset of 2017 data sample.
\section{NA62 Apparatus}
\label{sec:na62}
\begin{figure}[htb]
\centering
\includegraphics[width=0.9\textwidth]{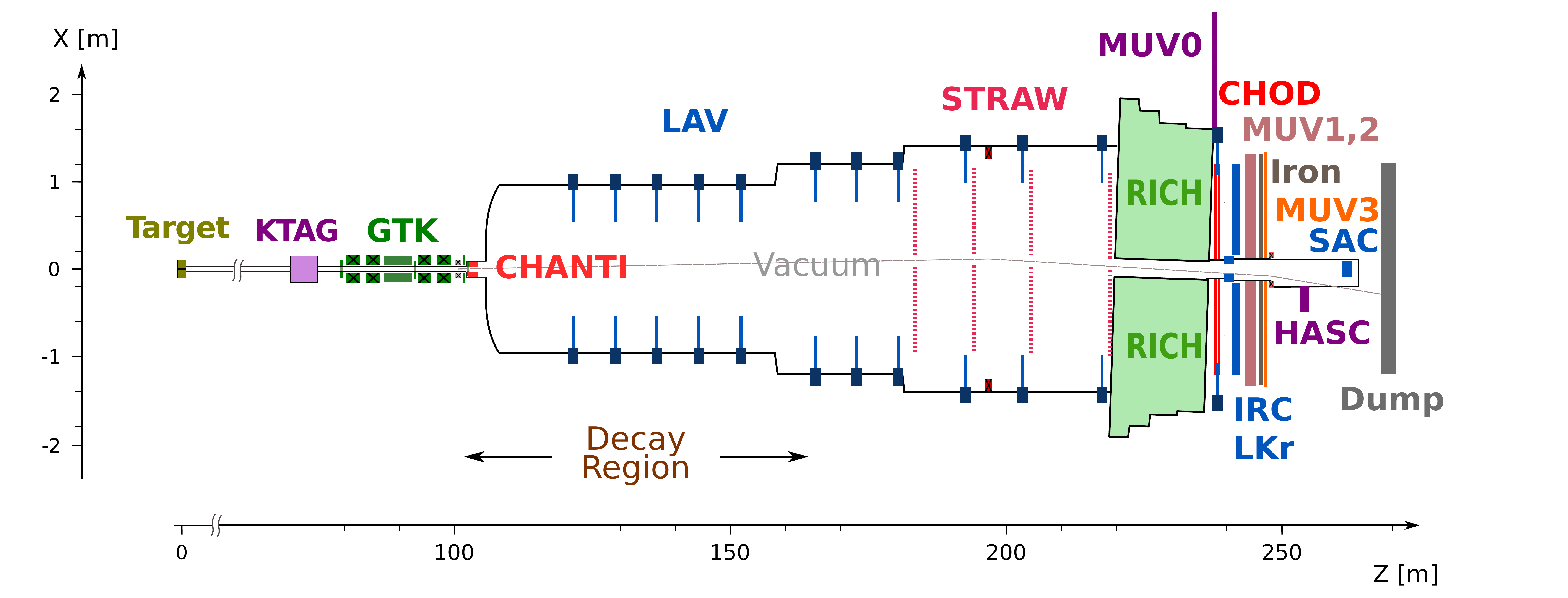}
\caption{\small Schematic xz view of the NA62 apparatus.}
\label{fig:na62}
\end{figure}

The fixed target NA62 experiment exploits a 400 GeV/c primary SPS proton beam.
A schematic view of its experimental apparatus is shown in Figure \ref{fig:na62}.
The proton beam impinges on a beryllium target producing a secondary beam, of which the kaon component is $6\%$.
A 100 m long beam line selects, collimates, focuses and transports positively charged particles of ($75.0 \pm 0.8$) GeV/c momentum
to the evacuated fiducial decay volume.
In order to select the kaon component of the secondary beam,
a differential Cherenkov detector filled with $N_2$, the KTAG, is used.
The KTAG provides also a precise time measurements for the kaons, while the tracking of them is entrusted to the Gigatracker (GTK) detector.
The GTK is composed of three silicon pixel stations of $6 \times 3$ cm$^2$ surface exposed to
the full 750 MHz beam rate, it
is used to timestamp and measure the momentum of the beam particles before entering the vacuum region.
The CHANTI detector, placed after the Gigatracker, tags hadronic beam-detector interactions in the last GTK station.
Downstream the decay region, a magnetic spectrometer (STRAW) is used to measure the momentum of the charged particles produced in the kaon decays. 
The STRAW is made of four straw chambers and a dipole magnet between the second and third chamber.
Two charged hodoscopes, CHOD and NA48-CHOD, are placed after the spectrometer, providing fast time response for charged particles.
The $K^+$ decay channels with largest BRs are $K^+\to \pi^+ \pi^0 (\gamma)$ and $K^+ \to \mu^+ \nu_{\mu} (\gamma)$.
For such reason hermetic photon veto and muon veto systems, alongside with a redundant particle identification system,
are needed.
The photon veto system is composed by the Large Angle Veto (LAV), which covers angles from 8.5 to 50 mrad
and a set of small angle vetoes (IRC, SAC, LKr) which covers up to 8.5 mrad.
The LKr is an electromagnetic calorimeter, 26 $X_0$ deep quasi-homogenous ionization
chamber, that allows to measure the full electromagnetic shower.
The muon veto system is composed by
two modules of iron-scintillator sandwiches (MUV1 and MUV2) 
and a fast scintillator array (MUV3) which identifies and triggers muons.
MUV1/2/3, together with the LKr constitute the calorimetric particle identification (PID) system,
which is complemented with the PID provided by a
Ring Imaging CHerenkov (RICH) detector.
The RICH, with a 17 m long tank filled with neon, distinguishes pions,
muons and positrons and measures the time of the passage of the decay products.
Details on the detectors and their performance is reported in \cite{ref:NA62TD}.
\section{$K^+ \to \pi^+ \nu \bar{\nu}$ analysis strategy}
\label{sec:pinunu}
The same $K^+ \to \pi^+ \nu \bar{\nu}$ ($\pi\nu\nu$ in the following) analysis approach is used both in 2016 and 2017 datasets,
with some improvements in 2017 dataset, mainly a better $\pi^0$ rejection.\\
\begin{wrapfigure}{r}{0.5\textwidth}
\centering
\includegraphics[width=0.48\textwidth]{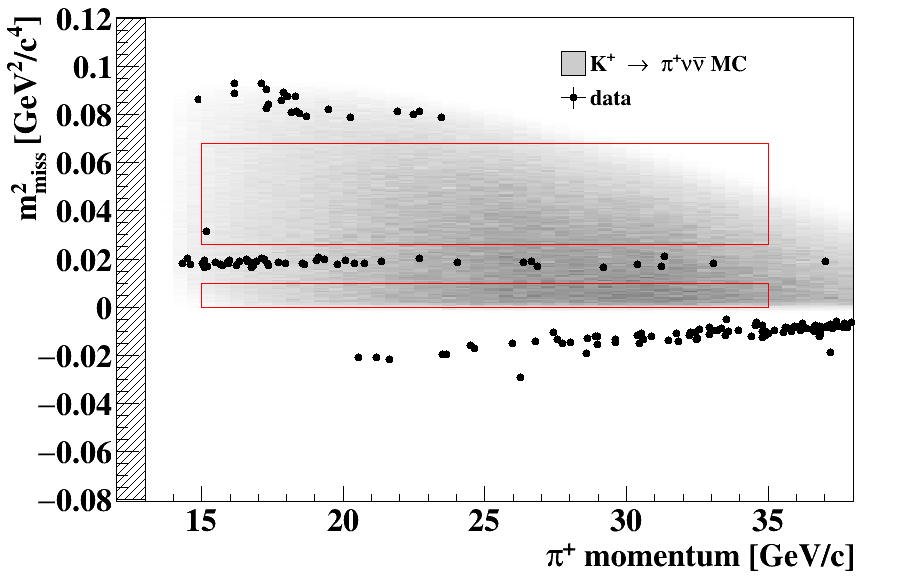}
\caption{Distribution of $m_{miss}^2$ versus $p_{\pi^+}$ for events passing the full event selection with 2016 dataset.}
\label{fig:pnn2016}
\end{wrapfigure}
The 2016 dataset analysis is described in \cite{ref:pnn2016} and proved the reliability and the working principle of the experimental technique.
The selected events are shown in Fig.\ref{fig:pnn2016}. 
The upper limit set was $BR(K^+ \to \pi^+ \nu \bar{\nu})< 14 \cdot 10^{-10}$ at $95\%$ C.L., corresponding to about 17 times the SM prediction.
In the following, a summary of the analysis strategy and preliminary sensitivity estimations are given for the 2017 dataset,
with a kaon sample 10 times larger than the 2016 one.  
The main ingredients for the $K^+ \to \pi^+ \nu \bar{\nu}$ analysis are the identification
and tracking of the particles before the decay region (upstream particles)
and the decay products (downstream particles).
Furthermore, the matching among them, through spacial and timing information, is a crucial aspect.
STRAW, RICH and CHOD are used to select a downstream particle.
If such a track is matched in time with a kaon candidate identified by the KTAG, 
an upstream track is searched for.
A discriminant built with the closest distance of approach and time differences between KTAG and GTK is used to choose the GTK track.
The decay vertex is reconstructed in the intersection of the
upstream and downstream tracks,
and is required to be in the fiducial volume (FV), which starts 10 m downstream the last GTK station and is 50 m long.
The events selected this way are defined as single-track events.
The key kinematic variable, which indicates the presence of neutrinos,
is the squared missing mass $m^{2}_{miss} = (P_{K^+}-P_{\pi^{+}})^{2}$,
where $P_{K^+}$ is the 4-momentum of the $K^{+}$ and $P_{\pi^+}$ is the 4-momentum of the charged
decay product under the $\pi^+$ mass hypothesis.\\

\begin{wrapfigure}{r}{0.6\textwidth}
\begin{center}
\includegraphics[width=0.58\textwidth]{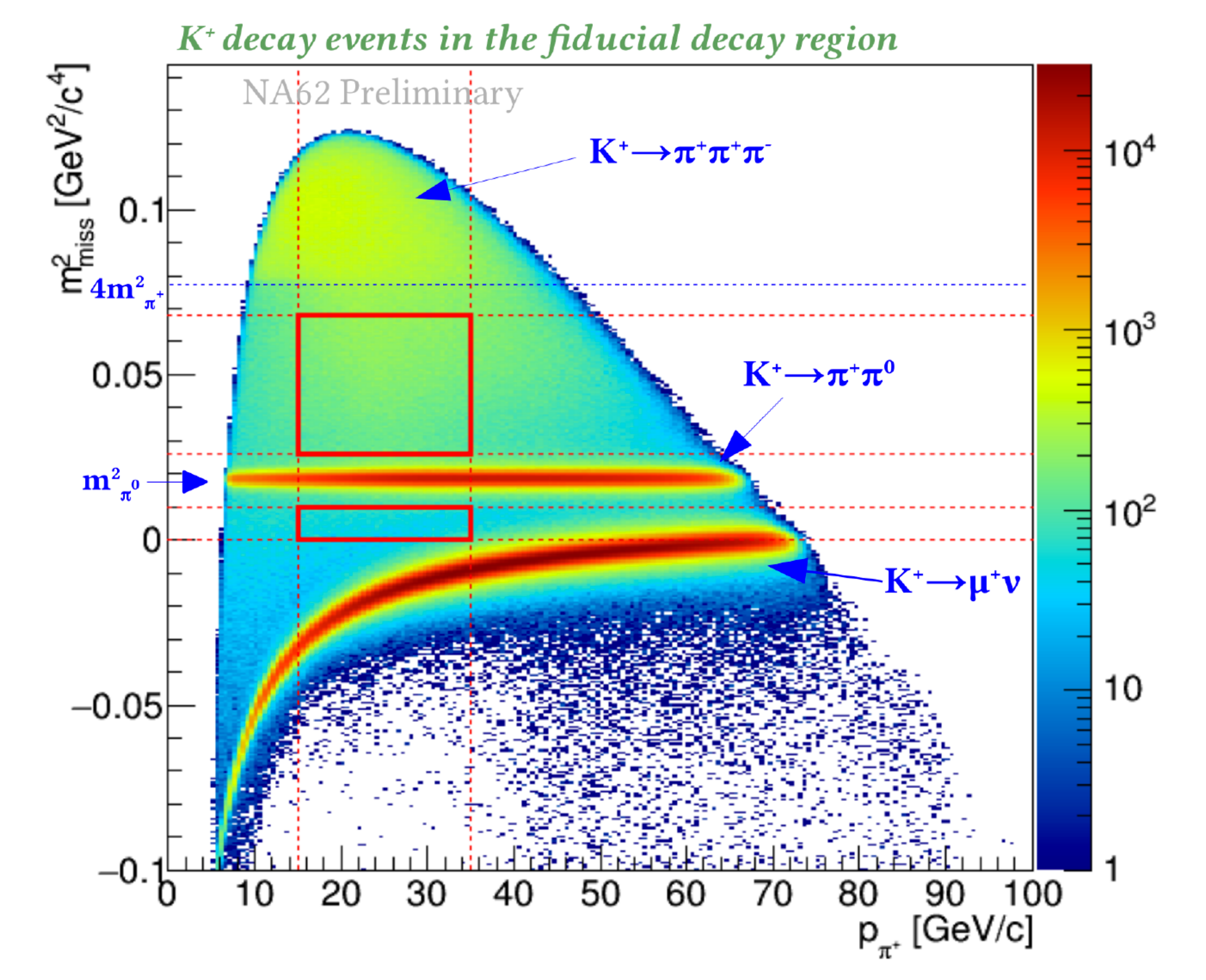}
\caption{Distribution of $m_{miss}^2$ versus $p_{\pi^+}$ for events passing the single-track event selection.}
\end{center}
\label{fig:pnn1trk}
\end{wrapfigure}
\begin{figure}[htb]
\centering
\includegraphics[width=0.95 \textwidth]{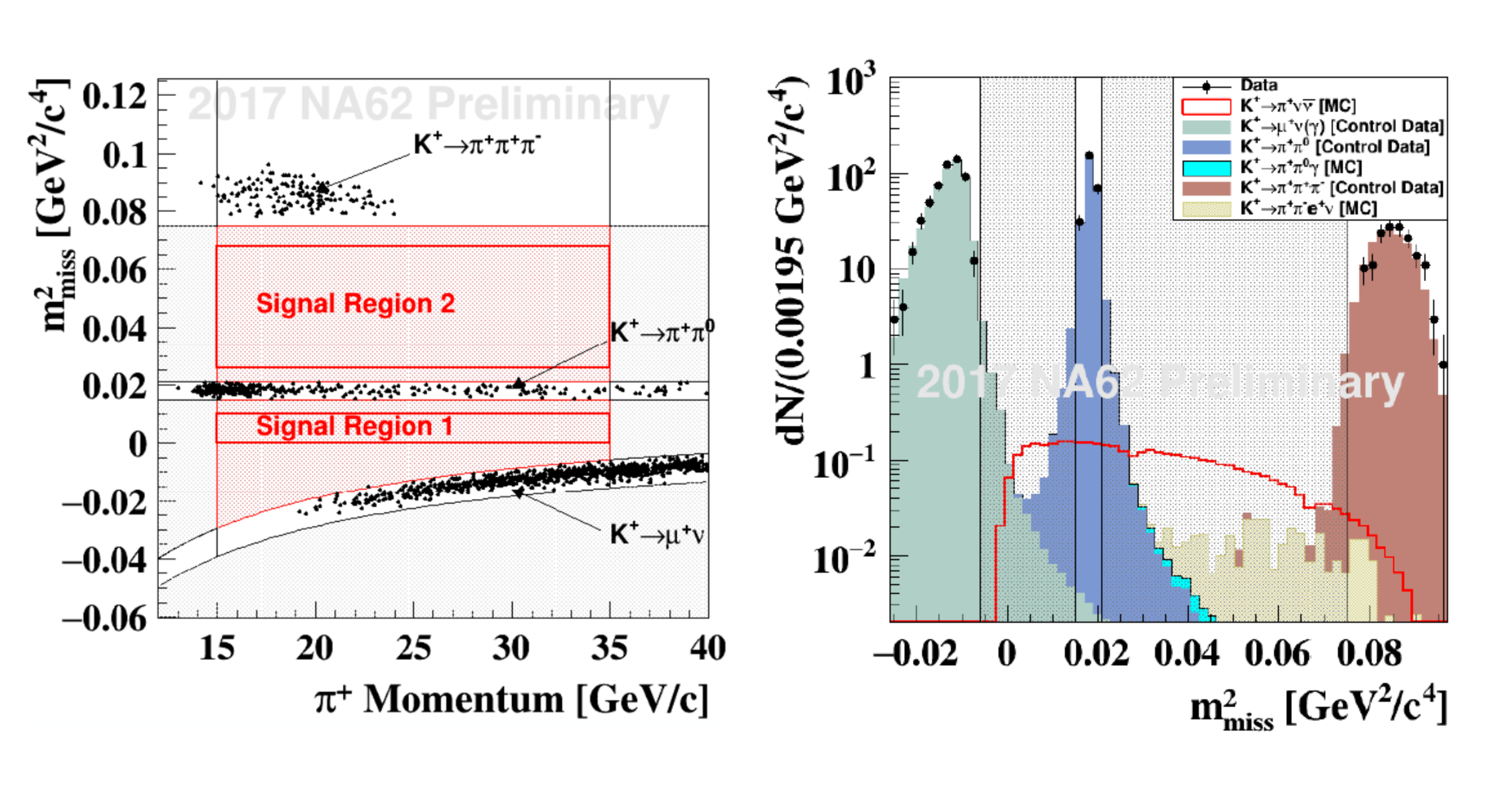}
\caption{ \small Left: Distribution of $m_{miss}^2$ versus $p_{\pi^+}$ for events passing the $\pi\nu\nu$ selection in the background regions.
  The control and signal regions are blinded. Right: Distribution of $m_{miss}^2$ for control samples and $\pi\nu\nu$
  sample only for the background $m_{miss}^2$ regions.} 
\label{fig:mmiss2}
\end{figure}

The $m^2_{miss}$ versus downstream track momentum distribution of the sample obtained with the single-track selection
on the 2017 dataset is shown
in Fig.\ref{fig:pnn1trk}, where the contributions of
the decays $K^+ \to \pi^+ \pi^0$, $K^+ \to \mu^+ \nu$ and $K^+\to \pi^+ \pi^+ \pi^-$ are well visible.
The pion identification is performed exploiting one algorithm using the calorimeter
system information (\textit{CaloPID}) and two algorithms employing the RICH PID.
For the CaloPID the energy deposits, energy sharing and shower shape profiles of the KLr, MUV1, MUV2,
together with signals in MUV3, are fed into a Boosted Decision Tree.
A likelihood based on the RICH measurements and STRAW tracking is built to give the probability
that a track corresponds to a $e^+,\mu^+$ or $\pi^+$ hypotheses.
In addition a cut on the mass reconstructed by the RICH, is applied \cite{ref:rich2018}.
These three algorithms give a total pion identification efficiency of
$64\%$ for a muon mis-identification probability of $10^{-8}$.
The photon rejection proceeds by combining information from LKr, LAV, SAC and IRC giving a
$\pi^0$ suppression of $\epsilon(\pi^0) = (1.4 \pm 0.1) \times 10^{-8}$.
A veto on multitrack events using CHOD signals is applied as well.
The momentum range considered is $[15,35]$ GeV/c.
Background, control and signal regions in $m^2_{miss}$ are defined considering the kinematic features
of the main background components 
and are indicated in Fig.\ref{fig:pnn1trk}.
The events which survived the full selection are shown in
Fig.\ref{fig:mmiss2} (left) for the background regions,
while the signal regions and the control regions were blinded at the time of the conference presentation.
The evaluation of the main background processes in the signal regions is performed with data-driven methods.
In particular, for the $K^+\to \pi^+ \pi^0$ and $K^+\to \mu^+ \nu$, control samples were
collected with a control trigger selected with  the $\pi\nu\nu$ criteria
except for the photon rejection (for $\pi^+ \pi^0$) and the pion identification (for $\mu\nu$).
These control samples are used to model the tails in the signal region.
These distributions, are normalized by the number of events, in the respective background regions,
after the full $\pi\nu\nu$ selection (including the $\pi^0$ rejection and the PID).
These two background contributions, together with other background kaon decays are shown in Fig.\ref{fig:mmiss2} (right).
The estimation of the upstream background, one of the most important contributions, was still under study during
the conference and is not included in the plot.
Preliminary sensitivity estimation presented at the conference and expected number of background events
is summarized in \cite{ref:spsc}, the Single Event Sensitivity (SES) is $SES = (0.34 \pm 0.04) 10^{-10}$,
which corresponds to an expected number of signal events in the SM
$N^{SM}_{\pi\nu\nu}= 2.5 \pm 0.4$.
 The signal expectation was normalized to $K^+\to \pi^+ \pi^0$
decays selected similarly to $K^+ \to \pi^+ \nu \bar{\nu}$, except for the
photon and multi-track rejection.
The background contamination from $K^+$ decays is $N_{bkg} = 0.76 \pm 0.10$, while the upstream background was still ongoing.  
The updated signal efficiency, normalization, complete background estimation and observed number of events in the signal
regions will be shown at the KAON2019 Conference (10 September 2019) and a CERN seminar (23 September 2019),
where preliminary results on the $BR(K^+ \to \pi^+ \nu \bar{\nu})$ will be given.
\section{Lepton Number Violation searches}
\label{sec:LNV}
A subset of 2017 dataset was used to perform searches for LNV processes, whose results are published in
\cite{ref:NA62LNV} and summarized in the following.
A dedicated trigger stream is used to collect three track final states.
Two channels have been considered: $K^+ \to \pi^- e^+ e^+$ and $K^+ \to \pi^- \mu^+ \mu^+$.
Both are normalized to their corresponding SM decay channels:
$K^+ \to \pi^- e^+ e^+$ is normalized to the SM decay $K^+ \to \pi^+ e^+ e^-$ and
$K^+ \to \pi^- \mu^+ \mu^+$ is normalized to the SM decay $K^+ \to \pi^+ \mu^+ \mu^-$.
That allows for cancellation of systematic uncertainties on efficiencies.
Three-track vertices are reconstructed by extrapolation of STRAW tracks upstream into
the FV and requiring that they are in time.
Particle identification is obtained with cuts on $E/p$, ratio of the energy deposit in LKr to the momentum measured by the STRAW.
The SM and LNV decays are selected in a range of the three body invariant mass $m_{\pi l l}$ around the kaon mass.
\subsection{$K^+ \to \pi^- \mu^+ \mu^+$ Analysis}
The track momentum range considered is 5 GeV/c $< p <$ 45 GeV/c.
The SM decay branching ratio is $BR(K^+ \to \pi^+ \mu^+ \mu^-) = (9.62 \pm 0.25) \times 10^{-8} $,
and the SM sample selected for the normalization has $m_{\pi \mu \mu}$ in the range:
$484$ MeV/c$^2 < m_{\pi \mu \mu} < 504 $ MeV/$c^2$.
The LNV signal region is defined as $|m_{\pi \mu \mu} - m_{K}|< 3 \delta  m_{\pi l l}$,
with  $\delta  m_{\pi \mu \mu}=1.1$ MeV/$c^2$ and was kept blinded till the freezing
of the selection and background evaluation.
The acceptance for the SM decay is evaluated with measured phase space density,
while the acceptance for the LNV decay is assumed to be uniform.
The main background contributions come from $K^+ \to \pi^+ \pi^+ \pi^-$ with  
in-flight pion decays and pion misidentified as a muon.
The number of expected total background events is $N_B = 0.91 \pm 0.41$.
After the unblinding 2 events are observed in the signal region, leading to an upper limit of
\begin{equation}
BR(K^+ \to \pi^- \mu^+ \mu^+)  < 4.2 \times 10^{-11}  \textrm{ at }  90 \%  \textrm{ C.L.}
\end{equation}
\begin{figure}[htb]
\centering
\includegraphics[width=0.4 \textwidth]{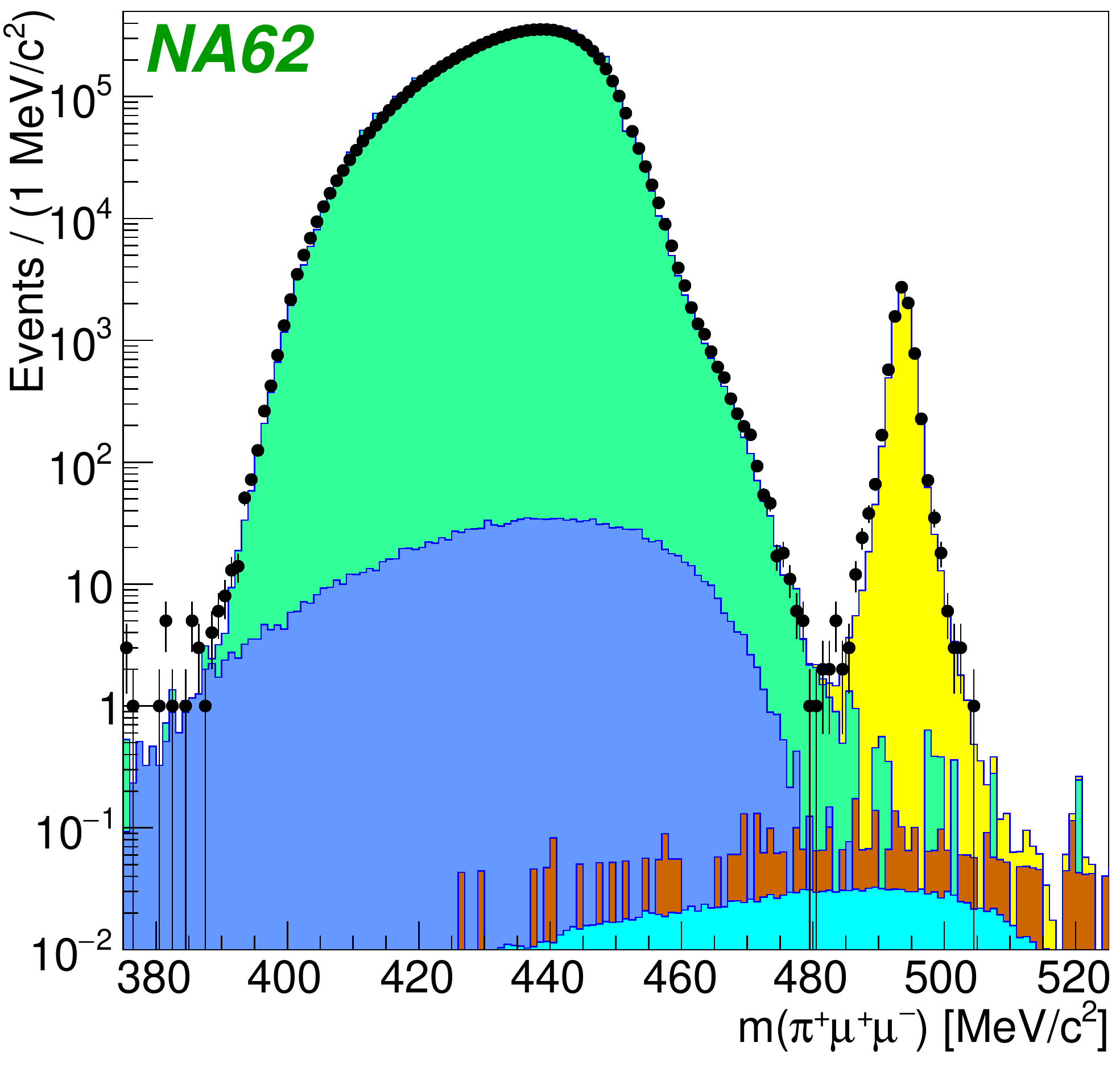} \hspace{1cm}
\includegraphics[width=0.4 \textwidth]{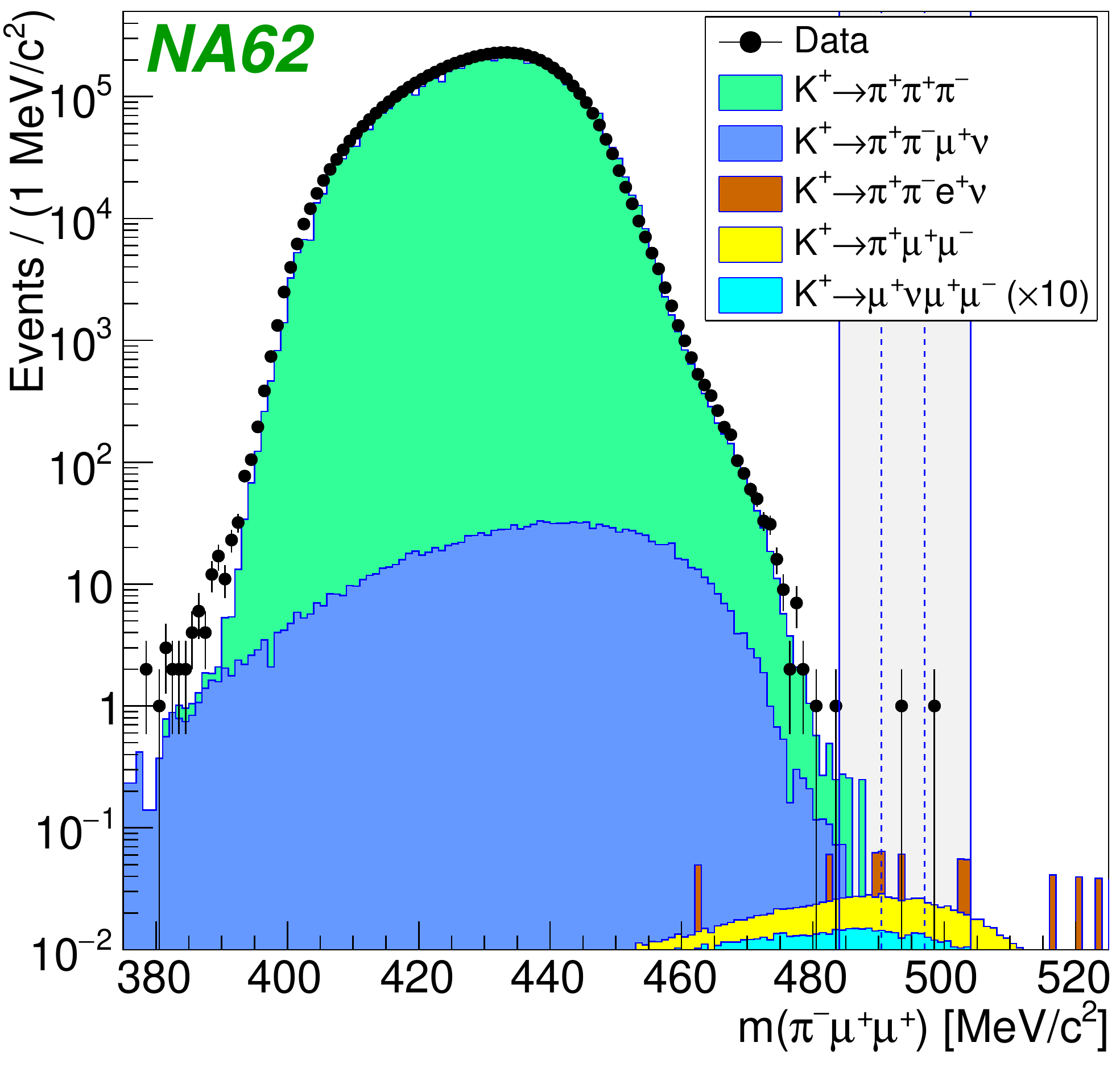}
\caption{\small Three-track invariant mass distribution for the $\mu$ channel for the SM decay (left) and the LNV decay (right).}
\label{fig:NA62}
\end{figure}

\subsection{$K^+ \to \pi^- e^+ e^+$ Analysis}
The track momentum range considered is 8 GeV/c $< p <$ 45 GeV/c.
Additional particle identification is applied to select $e^{+/-}$, using the track-based RICH likelihood algorithm.
In addition, LAV photon veto is applied to reduce the $\pi^0$ Dalits decays from $K^+\to \pi^+\pi^0$.
The SM decay branching ratio is $BR(K^+ \to \pi^+ e^+ e^-) = (3.00 \pm 0.09) \times 10^{-7} $.
The SM sample for the normalization, is selected in the $ m_{\pi e e}$ range: $484$ MeV/c$^2 < m_{\pi e e} < 504 $ MeV/$c^2$,
with the further request that $m_{ee} > 140$ MeV/$c^2$ to suppress
events with $\pi^0 \to e^+ e^-  \gamma$, $\pi^0 \to e^+ e^-  e^+ e^- $, $\pi^0 \to e^+ e^-$.  
The LNV signal region is defined as $|m_{\pi ee} - m_{K}|< 3 \cdot \delta  m_{\pi ee}$,
with  $\delta  m_{\pi e e}=1.7$ MeV/$c^2$.
The procedure for the acceptance estimation is analogous to the one for the $e$ channel.
The main background contribution is due to $K^+ \to \pi^0 e^+ \nu $, with $\pi^0 \to e^+ e^-$ and $e^-$ misidentified as a $\pi^-$.
The number of expected total background events is $N_B = 0.16 \pm 0.03$.
After the unblinding no events are observed in the signal region, leading to an upper limit of
\begin{equation}
  BR(K^+ \to \pi^- e^+ e^+)  < 2.2 \times 10^{-10}  \textrm{ at }  90 \%  \textrm{ C.L.,}
\end{equation}
  improving the previous upper limit of a factor of 2.

\begin{figure}[htb]
\centering
\includegraphics[width=0.4 \textwidth]{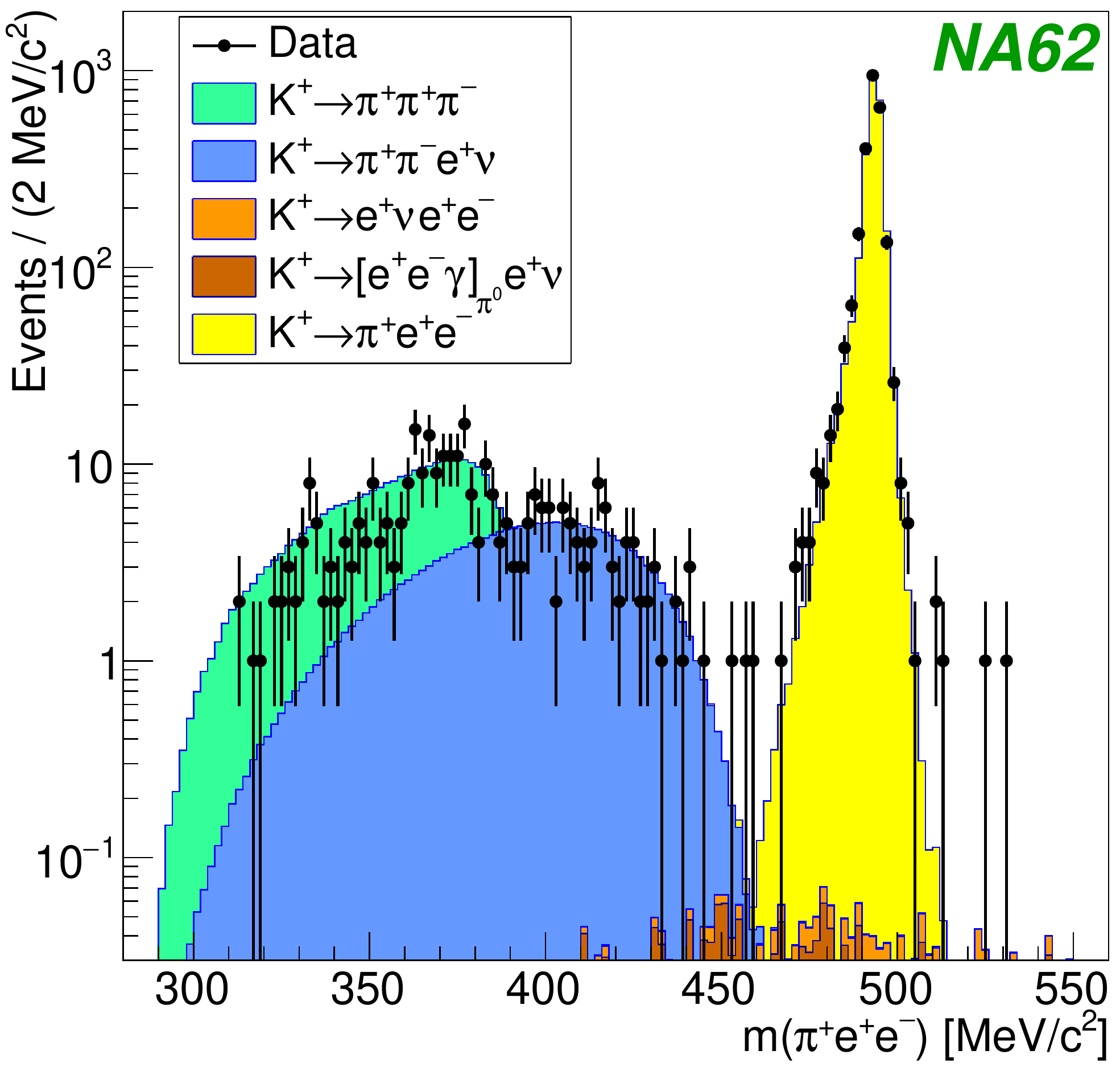} \hspace{1cm}
\includegraphics[width=0.4 \textwidth]{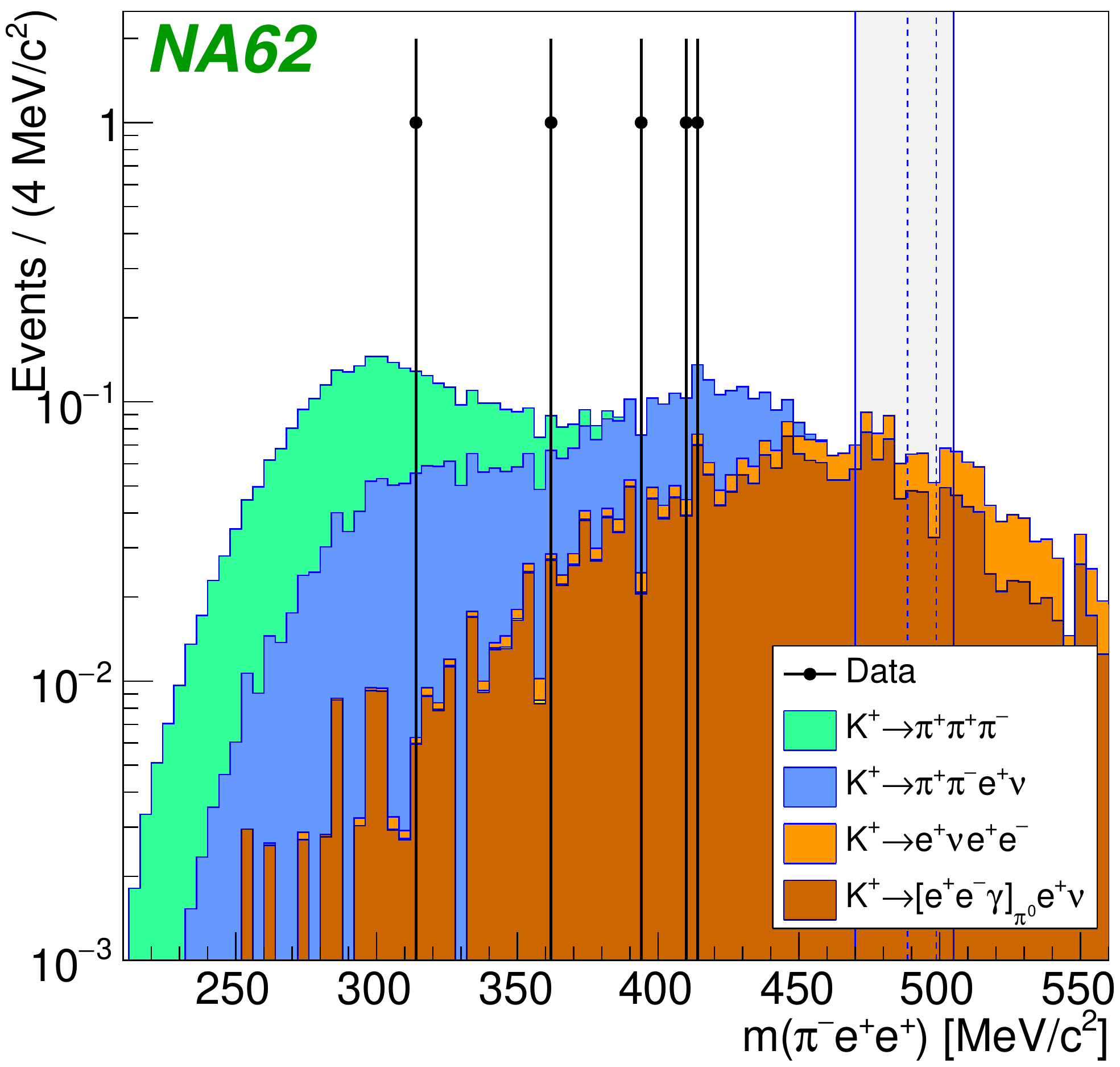}
\caption{\small Three-track invariant mass distribution for the $e$ channel for the SM decay (left) and the LNV decay (right).}
\label{fig:NA62}
\end{figure}
\section*{Acknowledgements}
R.Volpe was supported by FRS-FNRS under the Excellence of Science (EoS) project n.30820817
be.h \textit{The H boson gateway to physics beyond the Standard Model}.

\end{document}

%% file: econfmacros.tex
\def\beq{\begin{equation}}
\def\eeq#1{\label{#1}\end{equation}}
\def\eeqn{\end{equation}}

\def\beqa{\begin{eqnarray}}
\def\eeqa#1{\label{#1}\end{eqnarray}}
\def\eeqan{\end{eqnarray}}

\let\bar=\overbar

\def\Dslash{\not{\hbox{\kern-4pt $D$}}}
\def\dslash{\not{\hbox{\kern-2pt $\del$}}}

\def\msb{{\bar{\ssstyle M \kern -1pt S}}}

